# Thermally Configurable Multi-Order Polar Skyrmions in Multiferroic Oxide Superlattices

*Kefan Liu[1], Yuhui Huang[1, 2], Xiangwei Guo[1, *], Yongjun Wu[1, 2, 3, 4, *], Juan Li[5], Zijian Hong[1, 2, 3, 4, *]*

[1] State Key Laboratory of Silicon and Advanced Semiconductor Materials, School of Materials Science and Engineering, Zhejiang University, Hangzhou, Zhejiang 310058, China.

[2] Zhejiang Key Laboratory of Advanced Solid State Energy Storage Technology and Applications, Taizhou Institute of Zhejiang University, Taizhou, Zhejiang 318000, China.

[3] Institute of Fundamental and Transdisciplinary Research, Zhejiang University, Hangzhou 310058, China

[4] School of Engineering, Hangzhou City University, Hangzhou, Zhejiang 310015, China

[5] College of Materials Science and Engineering, Zhejiang University of Technology, Hangzhou 310014, China.

*Corresponding authors: l240441@zju.edu.cn (XG); yongjunwu@zju.edu.cn (YW); hongzijian100@zju.edu.cn (ZH)

**Abstract**

Polar topological textures in low-dimensional ferroelectrics have emerged as a versatile platform for high-density information storage and neuromorphic computing. While low-order topological states, such as vortices and skyrmions, have been extensively studied, high-order polar topological families remain largely unexplored due to their higher energy requirements and limited stabilization methods. Here, using a BiFeO$_3$ (BFO)-based multiferroic superlattice as a model system, we demonstrate a thermal-modulation strategy that stabilizes multi-order polar skyrmions and enables reversible tuning of their topological order through phase-field simulations. It was found that temperature modulation drives the system from polar solitons through 1$\pi$-, 2$\pi$-, 3$\pi$-, and 4$\pi$-skyrmion states, with closed heating-cooling path analyses revealing the widest thermal stability window for 2$\pi$-skyrmions (up to 600 K). Leveraging this robustness, 2% Sm doping in BFO lowers the transition temperatures, enabling room-temperature stabilization of 2$\pi$-skyrmions. These findings enrich the fundamental understanding of multi-order polar topologies and establish a tunable strategy for realizing variable-order topological configurations in practical memory devices.

# 1. Introduction

Topologically protected polar textures, which are nanoscale ferroic objects, have received growing attention due to their diverse physical properties and potential applications in advanced information technologies. [1–4] These applications include beyond-CMOS devices, racetrack memories, and neuromorphic computing systems. In recent years, various polar topological textures have been identified in low-dimensional ferroelectric heterostructures, such as polar flux closure, [5] vortices [6], merons [7], skyrmions [8], and Solomon rings [9], etc. These polar topologies provide a versatile platform for observing unconventional physical phenomena, resulting from the complex interactions between swirling polarization and external stimuli. A wide range of emergent properties has been reported, i.e., negative capacitance, [10,11] chirality, [12,13] ultrafast collective oscillations, [14] reversible inversion symmetry, [15] and ultrahigh energy-storage density, [16] offering exciting opportunities for fundamental physics research and the development of next-generation multifunctional nanoelectronics.

In addition to polar vortices, solitons, and skyrmions which have been widely observed, research has now shifted toward discovering and manipulating multi-order skyrmions ($k\pi$-skyrmion, $k$ = 1, 2, 3, ...) to expand the capacity of next-generation electronic applications. Multi-order skyrmions exhibit concentric azimuthal rotations of in-plane polarization around a central core, with their net topological charge $Q$ alternating between ±1 (for odd values of $k$, such as 1$\pi$- and 3$\pi$-) and 0 (for even values of $k$, such as 2$\pi$- and 4$\pi$-skyrmions) [17-21]. For example, a doughnut-like polar high-order radial vortex, analogous to magnetic skyrmionium, was recently observed in BiFeO$_3$ (BFO) nanoislands through boundary condition engineering [21]. This highlights both the feasibility of such structures and the current scarcity of related studies. However, it should be noted that multi-order skyrmions are inherently more challenging to stabilize and manipulate due to their larger spatial size and higher energetic cost [22-25]. The limited progress in this area underscores the urgent need for new strategies to explore, stabilize, and modulate multi-order polar topologies for future applications.

Thermally driven topological transitions have emerged as exciting and promising methods for manipulating polar topological states, allowing access to configurations that traditional methods cannot achieve. The change of temperature can induce deterministic switching between different textures by modulating the competition among temperature-dependent Landau energy, elastic energy, electrostatic energy, and gradient energy. [26, 27] Additionally, higher-order topologies are often higher-energy configurations which is often difficult to form naturally [28]. Controlled thermal excitation can, in principle, provide the necessary energy to overcome kinetic barriers and reveal otherwise inaccessible states. However, a key challenge lies in identifying a carefully tuned temperature window that can trigger topological transitions while remaining suitable for in-situ implementation. This would enable reversible thermal switching between multi-order polar topological structures and allow for precise control over their topological order.

In this study, we employ a BFO-based multiferroic superlattice as the model system and use a thermal control strategy to achieve stable multi-order polar skyrmion states. This approach enables reversible topological transitions among these states, allowing for precise control over their skyrmion order. By modulating temperature, we drive the system through successive skyrmion states, transitioning from $1\pi$ to $4\pi$ configurations. Our analysis of controlled heating and cooling paths demonstrates the reversibility of these multi-order transformations. Additionally, we predict that doping with Samarium (Sm) effectively lowers the transition temperatures and can even stabilize $2\pi$ skyrmions at room temperature for the superlattice system. This research enhances our fundamental understanding of multi-order polar topological states and establishes a pathway for realizing variable-order topological configurations in multistate memory devices.

## 2. Results and discussion
### 2.1. Diverse polar topologies in multiferroics and thermal modulation framework

The schematic diagrams illustrating various polar topologies, including vortices, solitons, and $k\pi$-skyrmions, are presented in **Figure 1(a)**. Among them, the high order skyrmions are usually high energy states that is difficult to stabilize at room temperature. In this study, using phase-field simulations (details in Methods) we propose a thermally driven modulation strategy to stabilize and manipulate the multi-order polar skyrmions. The $[(BiFeO_3)_7/(SrTiO_3)_4]_8$ superlattice (BFO/STO) grown on a $(001)_{pc}$-$LaAlO_3$ (LAO) substrate is chosen as the model system (**Figure 1b**), which has shown to host solitons at room temperature [29]. The spontaneous polarization state within the BFO layers is mainly influenced by the following key factors: the epitaxial constraints, which arise from the mechanical boundary conditions imposed by the LAO substrate; the depolarization field, resulting from the electrical boundary conditions introduced by the dielectric STO layers (**Figure 1c**); external heating, which affects the Landau energy landscape as the second-order Landau coefficient is modeled as a temperature-dependent tensor (details of this parameter can be found in **Supplementary Table S1**). [31–33] Further information about the simulation methodology is available in **Methods**.

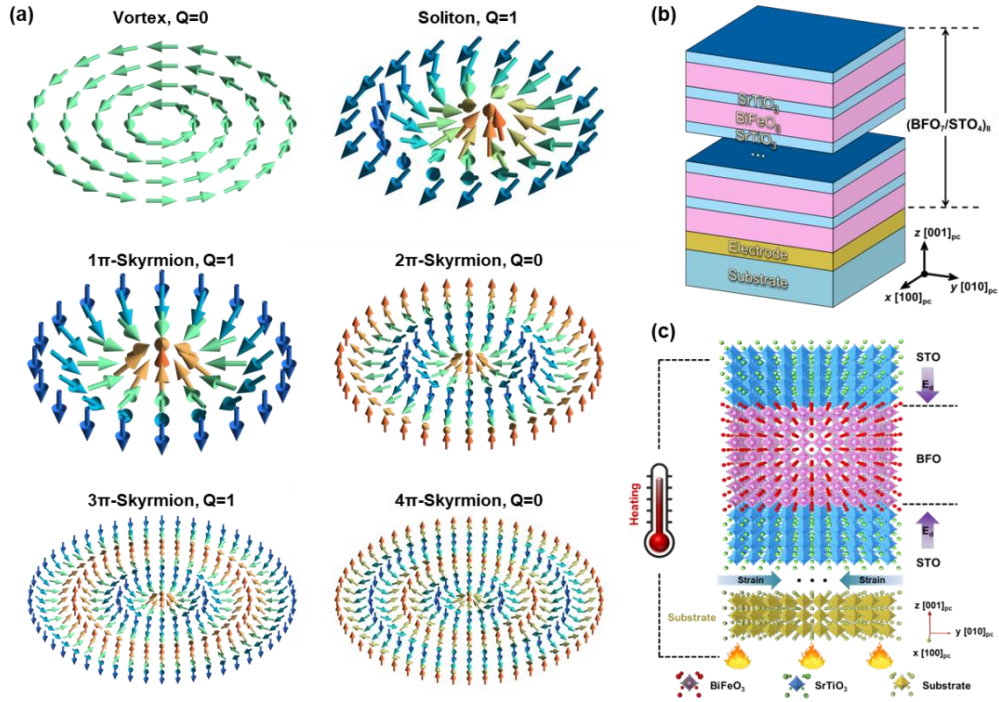

**Figure 1. Schematics of the topological polar textures and simulation setup.** (a) Representative polar textures characterized by different topological charge *Q*: polar vortex with *Q*=0, polar soliton with *Q*=1 and multi-order polar skyrmions with *Q*=0 or 1 (including 1π-, 2π-, 3π- and 4π-skyrmions). (b) Schematic of the phase-field simulation setup for the $BFO_7/STO_4$ superlattices. (c) Atomic schematic of the boundary conditions applied in the BFO/STO superlattices and influence of thermal field.

## 2.2. Thermally induced multi-order polar skyrmions in BFO/STO superlattices

The temperature-induced evolution of polar textures in the BFO/STO superlattices simulated by the phase-field simulations is presented in **Figure 2**. At room temperature (300 K), the system exhibits spatially dispersed solitons with a size of ~10 nm (**Figure 2a**) and a topological charge of 1 (calculated by integrating the local Pontryain density, **Figure 2b**), characterized by a bimeron-type texture with a converging in-plane polarization distribution (**Figure 2c**). The sparse distribution of soliton units gives rise to a relatively low-density topological phase. Upon heating to 600 K, a clear phase transition occurs wherein the polar solitons evolve into compact, regularly arranged low-order 1π-skyrmions (**Figure 2d,e**). This transformation is accompanied by the disappearance of bimeron structures and the emergence of symmetric skyrmionic textures (**Figure 2f**). Compared to the ~20 u.c.-wide polar solitons, the thermally stabilized 1π-skyrmions exhibit a reduced diameter of ~10 u.c. (**Figure 2g,h**), enabling a denser packing of topological units. Despite their smaller size, both topological states share a common polarization texture, featuring a downward-oriented core encircled by upward-pointing polarization, and a topological charge of $Q = -1$. This thermally driven polar soliton-to-1π-skyrmion transition consequently leads to a progressive increase in the total planar topological charge by 153% (**Figure 2i**), reflecting that thermal excitation can further enhance the potential information storage density to ~1.03 × $10^{13}$ bit/$in^2$.

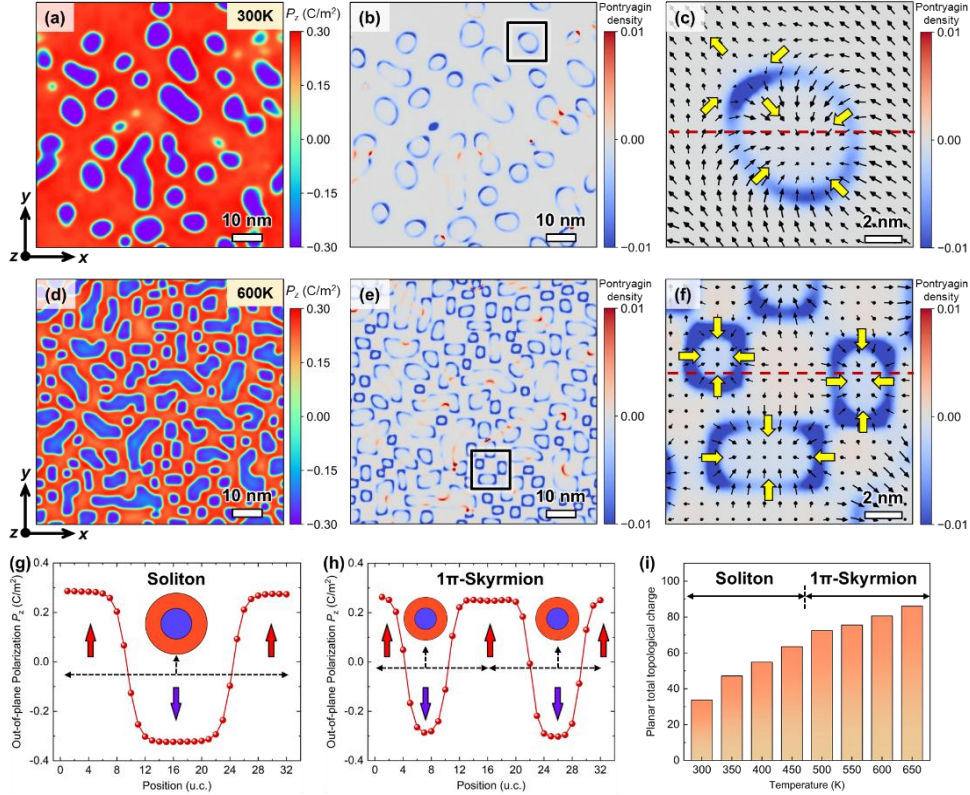

**Figure 2. The topological polar soliton at room temperature and emergent 1π-skyrmion under low thermal fields.** (a, b) Planar view of the out-of-plane polarization magnitude and the corresponding Pontryagin density distribution of BFO layer at 300K, showing the stabilization of polar soliton at the room temperature. (c) Magnified view of the polar soliton overlaid with the in-plane polar vector. (d, e) Planar view of the out-of-plane polarization magnitude and the corresponding Pontryagin density distribution of BFO layer at 600K, showing the formation of polar 1π-skyrmion under low thermal field. (f) Magnified view of the polar 1π-skyrmion overlaid with the in-plane polar vector. (g, h) Magnified view of the out-of-plane polarization distribution on a single soliton and 1π-skyrmion with red line profile overlaid. (i) Temperature dependence of the total topological charge of BFO layer during the transition from polar soliton to 1π-skyrmion induced by low thermal fields.

Interestingly, thermal excitation is further demonstrated to be effective in increasing the topological order of skyrmions under higher temperatures (**Figure 3**). At 800 K, a labyrinthine-type polar texture emerges in the system (**Figure 3a**), giving rise to a concentric nested skyrmion structure with opposing topological charges as revealed by Pontryagin density calculations (**Figure 3b**). Analysis of a single topological unit confirms a divergent polarization pattern ($Q = +1$) in the center and a convergent pattern ($Q = −1$) in the periphery, resulting in a net topological charge of zero (**Figure 3c**). Upon further heating to 1000 K, a clear labyrinth-to-bubble transition occurs, marked by the emergence of well-defined domain walls separating topological units (**Figure 3d**). Pontryagin density analysis identifies three concentric topological layers (**Figure 3e**), with an additional outer ring exhibiting a divergent polarization configuration, thereby increasing the net topological charge to +1 (**Figure 3f**).

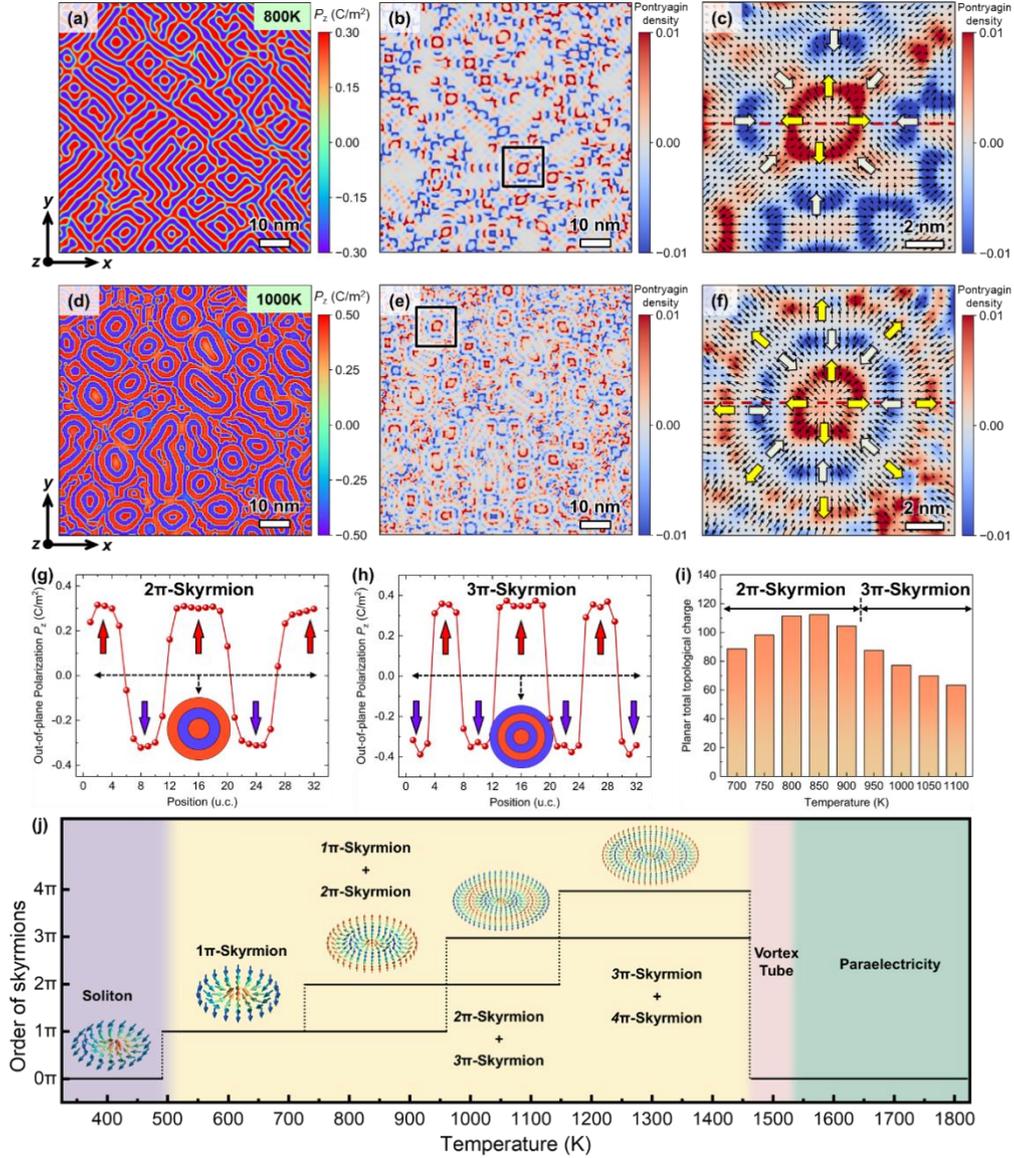

**Figure 3. Thermal driven phase transitions and temperature phase diagram for the BFO-based superlattice.** (a, b) Planar view of the out-of-plane polarization magnitude and the corresponding Pontryagin density distribution of BFO layer at 800K, showing the formation of mixed polar 1π-skyrmion and 2π-skyrmion under medium thermal fields. (c) Magnified view of the 2π-skyrmion overlaid with the in-plane polar vector. (d, e) Planar view of the out-of-plane polarization magnitude and the corresponding Pontryagin density distribution of BFO layer at 1000K, showing the formation of mixed polar 2π-skyrmion and 3π-skyrmion under medium thermal fields. (f) Magnified view of the polar 3π-skyrmion overlaid with the in-plane polar vector. (g, h) Magnified view of the out-of-plane polarization distribution on a single 2π-skyrmion and 3π-skyrmion with red line profile overlaid. (i) Temperature dependence of the total topological charge of BFO layer when the 2π-skyrmion and 3π-skyrmion were formed under medium thermal fields. (j) The topological temperature phase diagram with the order evolution of polar skyrmion.

The multi-order nature of these polar textures is further confirmed by analyzing the azimuthal rotation of out-of-plane polarization. At 600 K, a 2π rotation of the polarization is observed from the core to the outer region, corresponding to 2π-skyrmions (**Figure 3g**). While under a higher temperature of 1000 K, the polarization rotation extends to 3π, indicating the formation of 3π-skyrmions (**Figure 3h**). The larger diameter of 3π-skyrmions (~32 u.c.) compared to 2π ones (~24 u.c.) reduces their areal density, leading to a slight decrease in the overall topological charge (**Figure 3i**). This trend becomes more pronounced as the proportion of 3π-skyrmions increases. At 1400 K, the system evolves into four-layer nested textures, identified as 4π-skyrmions with zero net topological charge ($Q = 0$; **Supplementary Figure S1a-c**), accompanied by a further decline in total topological charge. A temperature-dependent topological order transition phase diagram was constructed for this process (**Figure 3j**). The system exhibits mixed-phase regions between 700-1400 K, where only two adjacent skyrmion orders (e.g., 1π and 2π) can coexist. The vortex-array phase is confined to a narrow window near the onset of the paraelectric transition.

### 2.3. Reversible thermal modulation of multi-order polar skyrmions

The above results demonstrate that heating effectively tunes the topological order of polar skyrmions. A key question for practical applications, however, is whether these multi-order topological states are thermally reversible upon cooling, and how robust their topological protection is under temperature cycling. To address this, we simulated cooling pathways by reversing the temperature sequence used during heating (**Figure 4a-d,f**). Remarkably, as the system cooled to 1000 K, 800 K, and 300 K, the corresponding 3π-skyrmions, 2π-skyrmions, and solitons were recovered in sequence, indicating an overall reversible evolution with only minor thermal hysteresis. However, at 600 K, a deviation from the heating path was observed: instead of 1π-skyrmions, 2π-skyrmions reappeared (**Figure 4e**). This behavior was reproduced across different cooling trajectories (**Supplementary Figure S2**), suggesting enhanced topological protection of 2π-skyrmions, which suppresses the reformation of the 1π-state. To test this further, reheating experiments were conducted (**Supplementary Figure S3**). When the system was first cooled to 400 K and then reheated to 600 K, it transitioned directly from solitons to 2π-skyrmions, bypassing the 1π-state. Only when cooled to 300 K and reheated to 600 K did the 1π-skyrmions re-emerge, confirming that the reappearance of the 1π-state depends sensitively on the cooling depth and path.

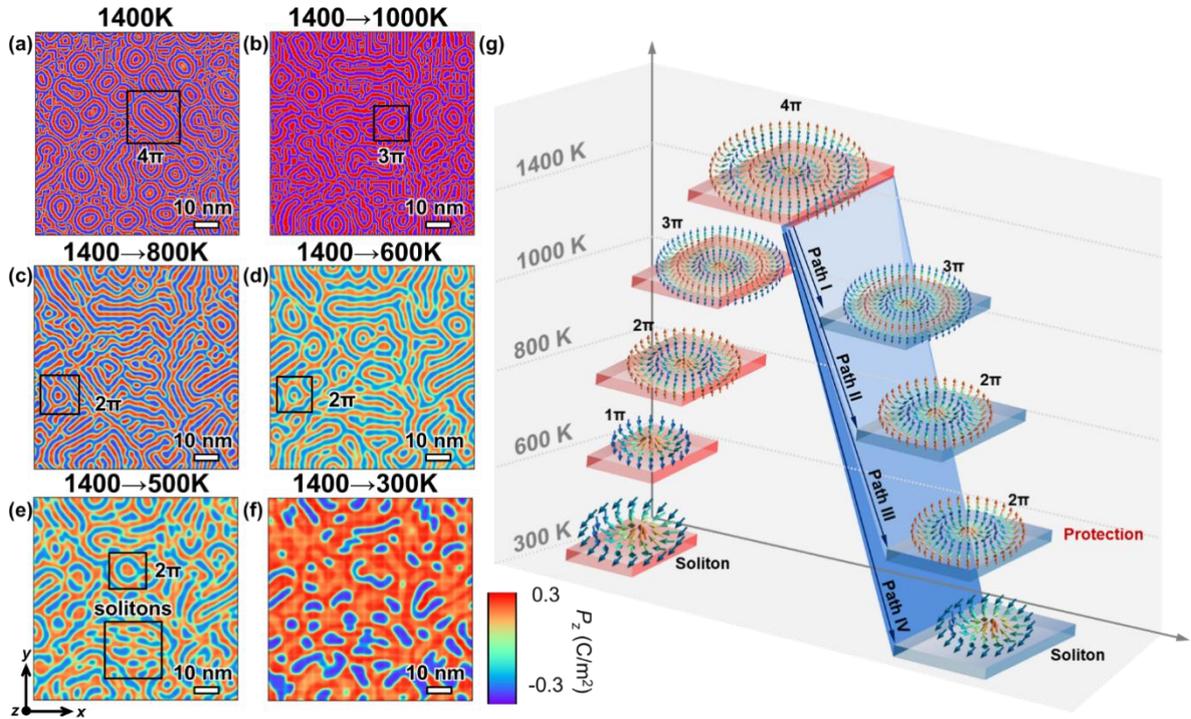

**Figure 4. Reversibility of multi-order polar topological structures during thermal modulation.** (a-f) Planar views of the out-of-plane polarization magnitude at different cooling stages: (a) 1400 K, (b) after cooling to 1000 K, (c) 800 K, (d) 600 K, (e) 500 K, and (f) 300 K. The evolution illustrates the transformation of multi-π polar textures, including their partial retention or decay into lower-order structures or solitons. (g) Schematic illustration of the temperature-driven transition pathways of multi-order polar topological phases, highlighting the reversibility and stability of distinct configurations.

### 2.4. Chemical doping-assisted stabilization of room-temperature multi-order skyrmions

Although thermal excitation can effectively induce multi-order skyrmions, the volatilization of $Bi_2O_3$ at elevated temperatures poses challenges for high temperature applications.[34–38] The wide thermal stability range of 2π-skyrmions, as reported above, nevertheless offers a promising alternative. To this end, we developed a strategy combining chemical doping, *i.e.*, via $Sm^{3+}$ doping in BFO (BSFO), with thermal pathways to stabilize 2π-skyrmions at room temperature (**Figure 5**). Sm doping could insert a chemical pressure that lowers both the Curie temperature and lattice constants, providing similar effect as thermal field (**Figure 5a**).[31–33] We first examined doping-induced topological transitions at room temperature by varying Sm content from 2% to 16% (**Supplementary Figure S4**). Increasing Sm concentration led to reduced topological unit size and increased density, yielding a transition from solitons to 1π-skyrmions. A Sm content-temperature phase diagram was constructed (**Figure 5b**). The phase diagram indicates that as Sm concentration increases, *k*π-skyrmions emerge at lower temperatures. However, as the Sm concentration increases, multi-order skyrmions gradually annihilate from higher to lower order. This is because the formation of multi-order skyrmions requires a larger initial volume of topological units. Sm doping reduces the size of topological units, thereby limiting the nucleation of multi-order skyrmions. It identifies an elliptical "2π-zone" where stable room-temperature 2π-skyrmions can form, particularly at 2% and 4% Sm concentrations. Further heating studies showed that only samples with ≤4% Sm content exhibited multi-order skyrmions (**Supplementary Figure S5-S8**), as the smaller topological units promoted by higher doping hinder the nucleation of complex structures.

Guided by this phase space, we applied the previously established thermal cycling protocol (**Figure 5c-e**). Comparative analysis of heating and cooling cycles for 2% and 4% Sm-doped samples (**Supplementary Figure S9-S10**) revealed that both could stabilize 2π-skyrmions at room temperature, though with different thermal requirements. Specifically, 4% Sm-doped BSFO required heating to 800 K, whereas 2% Sm-doped BSFO achieved similar skyrmion density at 600 K, an advantage for material integrity. Post-cooling observations showed that 2π-skyrmions evolved morphologically, with smoother peripheries and a shape transition from elliptical to bubble-like configurations (**Figure 5d,e**). Based on these findings, we propose an optimized protocol for room-temperature 2π-skyrmion generation (**Figure 5f**): (1) using 2% Sm-doped BFO, (2) thermal activation at 600 K, and (3) controlled cooling to room temperature. This three-step method reliably stabilizes 2π-skyrmions at operational conditions while minimizing thermal degradation.

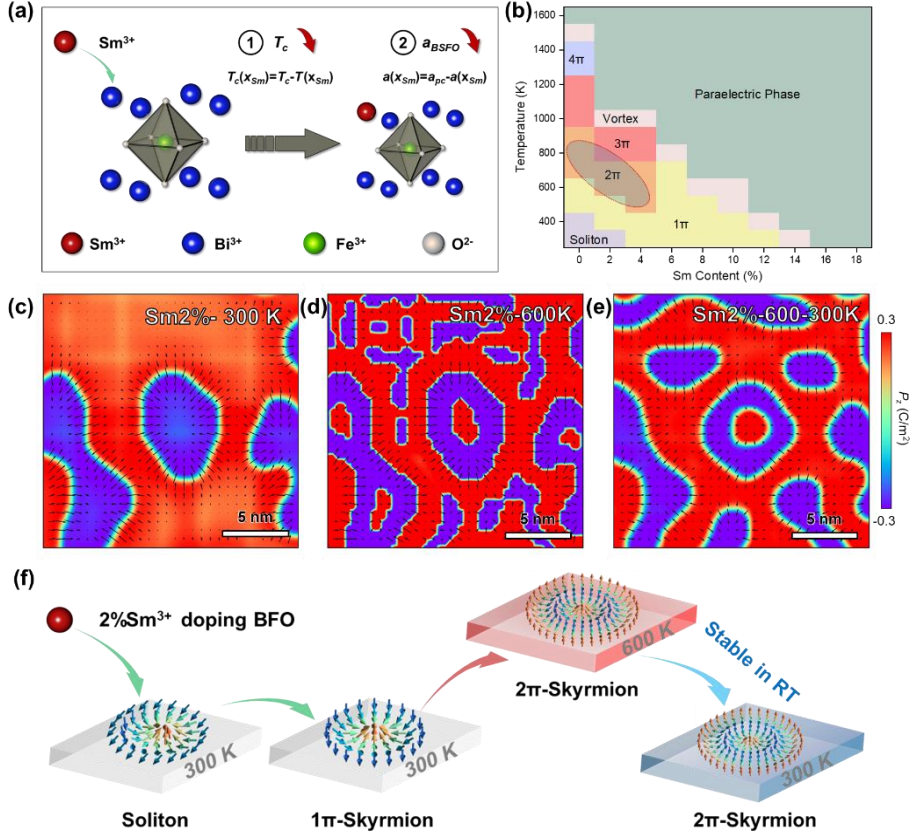

**Figure 5. Chemical doping combined with thermal path enables room-temperature stabilization of multi-order polar skyrmions.** (a) Schematic of the Sm-doping mechanism in BFO and simulation setup, illustrating the competition between Curie temperature modulation and epitaxial strain. (b) Sm concentration and temperature topological phase diagram. (c-e) Planar views of out-of-plane polarization magnitude overlaid with in-plane polarization vectors for 2% Sm doping at (c) 300 K, (d) 600 K, and (e) after cooling from 600 K to 300 K, revealing the thermal-path-dependent stabilization of 2π-skyrmions. (f) Schematic of room-temperature 2π-skyrmions enabled by Sm doping and controlled thermal evolution.

## 3. Conclusion and outlook

In summary, our phase-field simulations on BFO-based multiferroic superlattices demonstrate that thermal modulation offers an effective route to stabilize and control multi-order polar skyrmions. The system undergoes a sequential evolution from solitons to 1π-4π skyrmion states upon heating. Interestingly, it is noted that during the transition from solitons to 1π-skyrmions at 650 K, the density of topological units in the system increases by 153%, achieving an information storage density of $1.03 \times 10^{13}$ bit/in$^2$. The closed heating-cooling path analyses confirm reversible order switching of low- and high-order polar skyrmions, with 2π-skyrmions displaying the widest thermal stability window (up to 600 K). Furthermore, 2% Sm doping in BFO lowers the transition temperatures and, as evidenced by the Sm concentration-temperature phase diagram, defines an optimal regime for dense 2π-skyrmion formation, enabling their robust stabilization at room temperature. These findings not only highlight the power of thermal-control engineering as an effective strategy for constructing multi-order polar skyrmions in oxide superlattices but also enable customized doping protocols for multi-state memory systems, allowing precise stabilization of target polar topological states within specific operational temperature ranges.

## 4. Experimental Section

**Phase-field modeling of the BFO/STO superlattices**

Phase-field simulations are conducted to determine the equilibrium polarization configurations of the $(BFO_7/STO_4)_8$ superlattice epitaxially grown on LAO-$(001)_{pc}$, by numerically solving the time-dependent Ginzburg-Landau equation:[39–41]

$$\frac{\partial \boldsymbol{\xi}}{\partial t} = -L \frac{\delta F(\boldsymbol{\xi})}{\delta \boldsymbol{\xi}} \tag{1}$$

where $t$, $L$, $\boldsymbol{\xi}$ represent the evolution time step, the kinetic coefficient, and the order parameter, respectively. In this model, two feature vectors serve as order parameters: the spontaneous polarization vector ($\boldsymbol{P}$), and the oxygen octahedral tilt vector ($\boldsymbol{\theta}$). The total free energy $F$ is given by the volume integration of Landau ($f_{Land}$), elastic ($f_{elas}$), electrostatic ($f_{elec}$), and gradient energy densities ($f_{grad}$), which can be expressed by:

$$F(\boldsymbol{P}, \boldsymbol{\theta}) = \int \left[ f_{Land}(\boldsymbol{P}, \boldsymbol{\theta}) + f_{elas}(\boldsymbol{\varepsilon}, \boldsymbol{P}, \boldsymbol{\theta}) + f_{elec}(\boldsymbol{P}, \boldsymbol{E}) + f_{grad}(\boldsymbol{P}, \boldsymbol{\theta}) \right] dV \tag{2}$$

The Landau energy density is calculated using a fourth-order polynomial expansion:

$$f_{Land}(\boldsymbol{P}, \boldsymbol{\theta}) = \alpha_{ij} P_i P_j + \alpha_{ijkl} P_i P_j P_k P_l + + \beta_{ij} \theta_i \theta_j + \beta_{ijkl} \theta_i \theta_j \theta_k \theta_l + t_{ijkl} P_i P_j \theta_k \theta_l \tag{3}$$

where the $\alpha_{ij}$, $\alpha_{ijkl}$, $\beta_{ij}$, $\beta_{ijkl}$, and $t_{ijkl}$ are the Landau, oxygen octahedral tilt, and their coupling coefficients, respectively. The elastic energy density can be expressed as:

$$f_{elas}(\boldsymbol{\varepsilon}, \boldsymbol{P}, \boldsymbol{\theta}) = C_{ijkl}(\varepsilon_{ij} - \varepsilon_{ij}^0)(\varepsilon_{kl} - \varepsilon_{kl}^0) \tag{4}$$

where the $C_{ijkl}$ is the elastic stiffness coefficient tensor, and $\varepsilon_{ij}$ is the is the epitaxial strain applied to the substrate. The eigen strain $\varepsilon_{ij}^0$ is coupled to the spontaneous polarization and the oxygen octahedral tilt through:

$$\varepsilon_{ij}^0 = h_{ijkl} P_k P_l + \lambda_{ijkl} \theta_k \theta_l \tag{5}$$

where the $h_{ijkl}$ and $\lambda_{ijkl}$ are the electrostriction coefficient tensor. The $\varepsilon_{ij}$ can be determined by solving the elastic equilibrium equation ($\sigma_{ij,j} = 0$). The pseudocubic lattice parameters for BFO and STO are taken as 3.965 Å and 3.905 Å, respectively, for the lattice mismatch calculation. The elastic boundary conditions are defined such that the out-of-plane stress at the thin film surface vanishes, while the displacement at the substrate base far from the thin film is constrained to zero. The electric energy density can be performed as:

$$f_{elec}(\boldsymbol{P}, \boldsymbol{E}) = -\frac{1}{2}E_i P_i - k_{ij}\varepsilon_0 E_i E_j \tag{6}$$

where the $k_{ij}$ is the background dielectric constant, and the $\varepsilon_0$ is the vacuum dielectric constant. The calculation employs the electric equilibrium equation ($\sigma_{ij,j} = 0$), assuming that the equilibrium speed of the electric field is much faster than the domain evolution. The local electric field $E_i$ can be calculated by $E_i = -\nabla_i \varphi$. The electric boundary conditions are set where the electric potential $\varphi$ is zero at both the film-substrate interface and the top surface of the film. The gradient energy density can be calculated as:

$$f_{grad}(\boldsymbol{P}, \boldsymbol{\theta}) = g_{ijkl}\nabla_j P_i \nabla_l P_k + \kappa_{ijkl}\nabla_j \theta_i \nabla_l \theta_k \tag{7}$$

where the $g_{ijkl}$ and $\kappa_{ijkl}$ are the polar and rotation gradient coefficient, respectively.

A three-dimensional computational domain was discretized using a mesh of 200 × 200 × 150 grids, with a uniform grid spacing of 0.4 nm. Along the out-of-plane direction, the grid distribution was allocated as follows: 30 grids for the substrate, 87 grids for the superlattice, and 33 grids for the air layer. To simulate initial polarization nucleation, random noise with a small amplitude (<0.0001 μC·cm$^{-2}$) was introduced into the system. The temperature was uniformly applied to the entire system. For the heating process, we used the room-temperature results as initial configurations for the heating simulations to ensure consistency with actual experiments.

**Topological density calculation**

The topological charge is characterized by integrating the Pontryagin density ($q$) of thin film horizontal cross-sectional slices, calculated by:[7]

$$Q = \iint q \, dxdy = \iint \frac{1}{4\pi} \boldsymbol{P} \cdot \left(\frac{\partial \boldsymbol{P}}{\partial x} \times \frac{\partial \boldsymbol{P}}{\partial y}\right) dxdy \tag{8}$$

**Sm doping of BFO layer**

To account for the Sm doping effects, modifications were applied to the second-order Landau coefficients:[31]

$$\alpha_{11} = \alpha_0[T - T_c(x)] = \alpha_0[T - (T_c - c_p x)] \tag{9}$$

$$\beta_{11} = \beta_0[T - T_q(x)] = \beta_0[T - (T_q - c_q x)] \tag{10}$$

where $\alpha_{11}$ and $\beta_{11}$ represent the modified Landau coefficients for polarization and oxygen octahedral rotation, respectively. $T_c$ denotes the Curie temperature. $\alpha_0$, $\beta_0$, $c_p$, and $c_q$ are material-specific constants (see **Supplementary Table S1** for complete parameter values). The $x$ means the content of doped Sm.

Additionally, the compressive strain in the superlattice decreases with increasing Sm content. The epitaxial strain in the superlattice is modified with Sm content according to the following equation:

$$\bar{\varepsilon}_{11}(x) = \bar{\varepsilon}_{22}(x) = \frac{[a_s - a_f(x)]}{a_s} \quad (11)$$

where $a_s$ and $a_f$ represent the lattice constant of the substrate and film, respectively. Here we treat both the substrate and thin film as pseudocubic structures. The details of Sm doping-induced variations in BSFO lattice constants are shown in **Supplementary Figure S11**.[33]

**Acknowledgements**

Z. H. and K. L. are grateful for the technical support for Nano-X from Suzhou Institute of Nano-Tech and Nano-Bionics, Chinese Academy of Sciences (SINANO). The financial supports from the National Natural Science Foundation of China (Nos. 92166104, 92463306, 12174328, ZH), the Joint Funds of the National Natural Science Foundation of China (No. U21A2067, YW), and the Natural Science Foundation of Zhejiang Province (No. LR25E020003, ZH; No. LD24E020003, YW) are acknowledged. X.G. is supported by the National Natural Science Foundation of China (No. 52202151) and China Postdoctoral Science Foundation (No. 2022M722715). YW is also supported by the Special Support Plan for High Level Talents in Zhejiang Province (No. 2023R5231).

**Conflict of Interest**

The authors declare no competing interests.

**Data Availability Statement**

All data used are available within this manuscript and Supplementary Information. Further information can be acquired from the corresponding authors upon reasonable request.

**Code Availability Statement**

The phase-field simulation was performed with the Mu-PRO software package (https://muprosoftware.com).

**Author contributions**

Kefan Liu: Conceptualization, Data curation, Formal analysis, Visualization, Writing–original draft, Writing–review and editing.

Yuhui Huang: Writing–review and editing, Validation, Supervision.

Yongjun Wu: Writing – original draft, Writing – review and editing, Supervision, Funding acquisition, Resources.

Xiangwei Guo: Conceptualization, Methodology, Software, Writing–review and editing.

Juan Li: Funding acquisition, Resources, Writing–review and editing.

Zijian Hong: Conceptualization, Writing–original draft, Writing–review and editing, Formal analysis, Funding acquisition, Resources, Software, Validation .

**Supporting Information**

Supporting Information is available from the corresponding author (ZH).

**TOC figure**

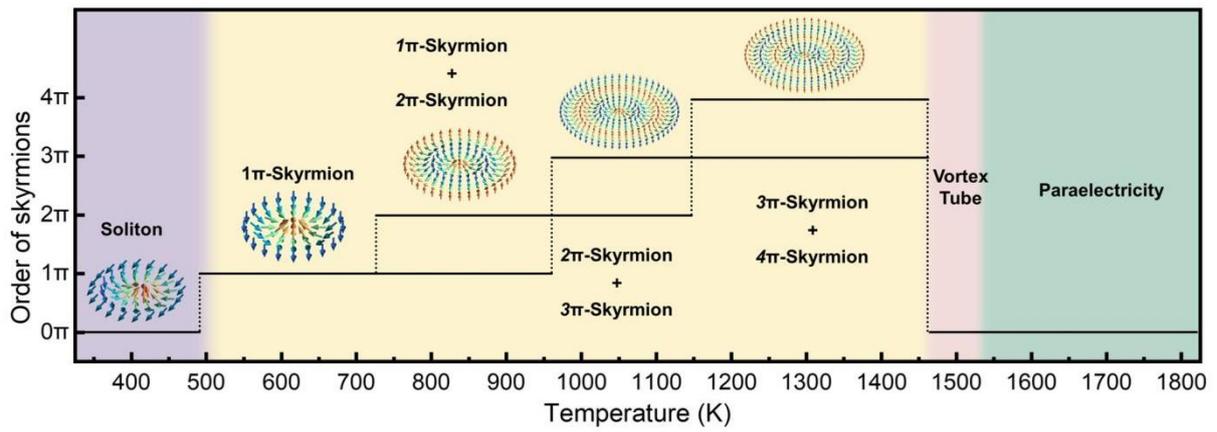

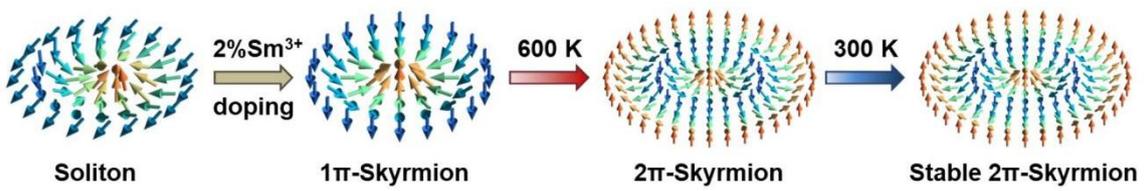